\documentclass[aip,twocolumn,apl,floatfix,superscriptaddress,showpacs,lengthcheck,hyperref,citeautoscript,reprint]{revtex4-1}
\usepackage{amsmath}
\usepackage{amssymb}
\usepackage{graphicx}

\usepackage{epsf}
\usepackage{epsfig}
\usepackage{braket}
\usepackage{bm}
\usepackage{amsfonts}
\usepackage{color,soul}

\newcommand{\kt}{k_{\text{B}}T}
\newcommand{\bea}{\begin{eqnarray}}
\newcommand{\eea}{\end{eqnarray}}

\begin{document}

\title{Diffusion of fluorine adatoms on doped graphene}
\author{R. M. Guzm\'an-Arellano}
\affiliation{Centro At{\'{o}}mico Bariloche and Instituto Balseiro, CNEA, 8400 Bariloche, Argentina}
\affiliation{Consejo Nacional de Investigaciones Cient\'{\i}ficas y T\'ecnicas (CONICET), Argentina}
\author{A. D. Hern\'andez-Nieves}
\affiliation{Centro At{\'{o}}mico Bariloche and Instituto Balseiro, CNEA, 8400 Bariloche, Argentina}
\affiliation{Consejo Nacional de Investigaciones Cient\'{\i}ficas y T\'ecnicas (CONICET), Argentina}
\author{C. A. Balseiro}
\affiliation{Centro At{\'{o}}mico Bariloche and Instituto Balseiro, CNEA, 8400 Bariloche, Argentina}
\affiliation{Consejo Nacional de Investigaciones Cient\'{\i}ficas y T\'ecnicas (CONICET), Argentina}
\author{Gonzalo Usaj}
\email[Corresponding author:]{usaj@cab.cnea.gov.ar}
\affiliation{Centro At{\'{o}}mico Bariloche and Instituto Balseiro, CNEA, 8400 Bariloche, Argentina}
\affiliation{Consejo Nacional de Investigaciones Cient\'{\i}ficas y T\'ecnicas (CONICET), Argentina}
\date{\today}

\begin{abstract}
We calculate the diffusion barrier of fluorine adatoms on doped graphene in the diluted limit using Density Functional Theory. We found that the barrier $\Delta$ strongly depends on the magnitude and character of the graphene's doping ($\delta n$): it increases for hole doping ($\delta n<0$) and decreases for electron doping ($\delta n>0$). Near the neutrality point the functional dependence can be approximately by   $\Delta=\Delta_0-\alpha\, \delta n$
 where $\alpha\simeq6\times10^{-12}$ meVcm$^2$. This effect leads  to significant changes of the diffusion constant with doping even at room temperature and could also affect the low temperature diffusion dynamics due to the presence of substrate induced charge puddles. In addition, this might open up the possibility to engineer the F dynamics on graphene by using local gates.
\end{abstract}




\maketitle

The physical properties of adatoms on graphene are the subject of intense research activity motivated by the possibility of tuning graphene electronic and magnetic properties by adding small impurity concentrations. Interesting examples include magnetic impurities,\cite{Uchoa2008,Yazyev2007,Palacios2008,Yazyev2010,Sofo2012} that might lead to Kondo physics,\cite{Vojta2004,Wehling2010,Cornaglia2009} or  spin relaxation,\cite{Tombros2007,CastroNeto2009,Han2011,Kochan2014} impurity induced localization,\cite{Ostrovsky2006,Gattenloehner2013,Roche2012,Hong2011,Usaj2014,Cresti2013} and many others.\cite{Yazyev2010,DasSarma2011} 
Determining the way different adatoms or molecules attach to graphene is at the core of many studies\cite{Duplock2004,Meyer2008,Chan2008,Wehling2009,Wehling2010a,Boukhvalov2009,Ao2010} as this defines the way they affect the otherwise planar structure of the pristine graphene. The charge transfer from the impurity orbitals to the host graphene, the formation of magnetic structures, clustering and diffusion of adatoms\cite{Lehtinen2003,Wehling2009} are properties determined by the nature of the chemical bonding of the adsorbates.

In the case of fluorine atoms, it was shown\cite{Sofo2011} that graphene's doping can play a major role in the way they link to graphene. In fact, a covalent-ionic like transition was predicted to occurs as electron doping is increased. Later, it was found that other types of adatoms may behave in similar ways.\cite{Chan2011}
This change in the bonding leads to a modification of the graphene's local structure, going from an sp$^3$-like coordination to an sp$^2$ one (see below), which is expected to strongly affect the electron scattering mechanism (both charge and spin).\cite{DasSarma2011,CastroNeto2009} This change is due to a subtle competition of the many contributions to the total energy of the system.\cite{Sofo2011} Evaluating total energies in relaxed structures requires the use of first principle calculations. 

Here we use Density Functional Theory (DFT) to estimate the dependence on electron and hole doping of the total energy of fluorine adatoms in different positions. Namely, we analyze the change in the diffusion barrier of fluorine on graphene induced by doping. Indeed, as might be expected from Ref. [\onlinecite{Sofo2011}], our results show that fluorine diffusion is strongly affected by the amount and type of doping. While this is similar to what happens in the case of oxygen,\cite{Suarez2011} here the diffusion barrier in the neutral case is much smaller, and hence our starting point correspond to a faster diffusive dynamics. In addition, in our case the change of the barrier occurs at lower doping levels. This opens the possibility to engineer the dynamical properties of F (and possible other adatoms\cite{Chan2011}) by local gating, freezing or speeding up the diffusion. 

\begin{figure}[b]
\includegraphics[width=.9\columnwidth]{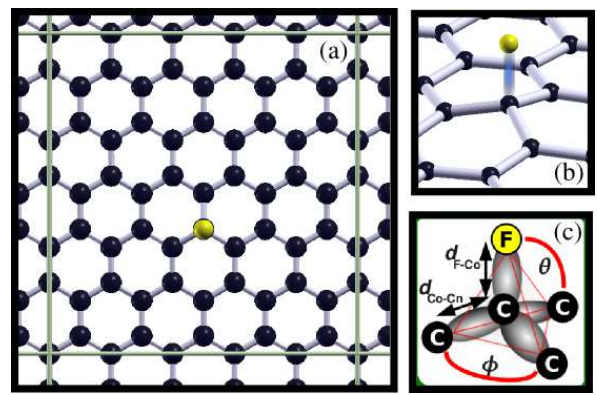}
\caption{(color online) (a) Unit cell used in the DFT calculation containing $60$C atoms and a $F$ atom. (b) Local  distortion of the graphene lattice below the adatom. (c) Definition of the geometrical parameters that define the F-C$_0$ bond. Carbon atoms are represented by black spheres and the fluorine atom by a yellow (light gray) sphere. }
\label{geometry}
\end{figure}

The DFT calculations were performed with the Quantum Espresso package \cite{QE} employing density functional theory and the Perdew-Burke-Ernzerhoff (PBE) exchange-correlation functional.\cite{pbe} A PAW description of the ion-electron interaction \cite{Blochl1994} was used together with a plane-wave basis set for the electronic wave functions and the charge density, with energy cutoffs of $70$ and $420$ Ry respectively. The electronic Brillouin zone integration was sampled with an uniform k-point mesh ($2\!\times\! 2\!\times\! 1$ or $4\!\times\! 4\!\times\! 1$ depending on the size of the supercell---$60$ and $32$ C atoms, respectively) and 
a Gaussian smearing of $0.01$ Ry. The two-dimensional behavior of graphene was simulated 
by adding a vacuum region of $20$ \AA~ above it. All the structures were relaxed using a criteria 
of forces and stresses on atoms of $0.005$eV/\AA~ and $0.5$GPa, respectively. The convergence tolerance of energy was set to $10^{-5}$ Ha ($1$ Ha = $27.21$ eV). To correct for the dipole moment generated in the cell and to improve convergence with respect to the periodic cell size, monopole and dipole corrections were considered.\cite{Neugebauer1992} This is particularly important in the doped cases. Doping  of the unit cell (added/removed  electrons) where compensated by an uniformly distributed background charge.
The diffusion barriers were calculated using the Nudge Elastic Bands method (NEB) as implemented in the QE package.  

Figure \ref{geometry} shows the unit cell geometry  used for our DFT calculations (containing $60$C atoms, unless otherwise specified) as well as the local geometry of the F-graphene bonding: the F atom sits on top of a C atom (C$_0$) forming a covalent bonding with it 
(undoped case). The bonding  corresponds to a distorted local sp$ ^3$-like hybridization of the C$_0$ atom. Hence, the $\theta$ and $\phi$ angles of the bond, defined in the figure, have values close to the ideal tetrahedral case ($\sim\!109.5^\circ$). The C$_0$ atom is slightly above the graphene sheet, roughly $0.5$\AA, in order to satisfy the local symmetry of the sp$ ^3$-like bond.

\begin{figure}[tb]
\includegraphics[width=.85\columnwidth]{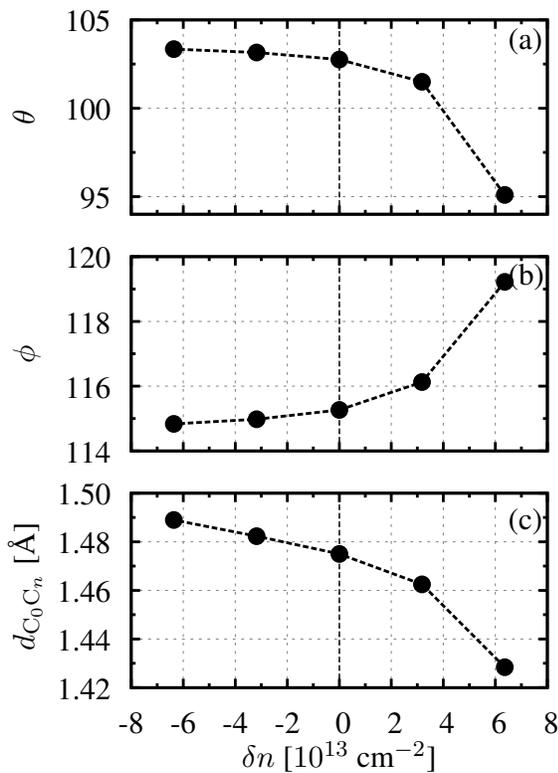}
\caption{Doping dependence of the F-C$_0$ bond parameters (see Fig. \ref{geometry} for their definition).  $\delta n >0$ ($\delta n <0$) indicates electron (hole) doping. The dots correspond to a change of $-1,-\frac{1}{2},0,\frac{1}{2},1$ electrons per unit cell.\cite{notes1}
The transition from a sp$^3$-like to a sp$^2$-like character of the bond is apparent from the figure.}
\label{angles}
\end{figure}

The change of the local geometry upon changing the graphene doping, $\delta n$, is presented in Fig.~\ref{angles} where $\delta n=0$ corresponds to the undoped case. Quite clearly, we observe that the systems undergoes a  sp$^3$ $\rightarrow$ sp$^2$ type transition\cite{Sofo2011} when we move from the hole ($\delta n<0$) to the electron  ($\delta n>0$) doped case. Namely, $\theta$ goes from $\sim\!103^\circ$ to $\sim\!95^\circ$ and $\phi$  from $\sim\!115^\circ$ to $\sim\!119^\circ$. Notice that also the distance between C$_0$ and its nearest neighbors  C-atoms becomes closer to the bulk C-C distance in graphene ($a_0=1.42$\AA). 

\begin{figure}[tb]
\includegraphics[width=.95\columnwidth]{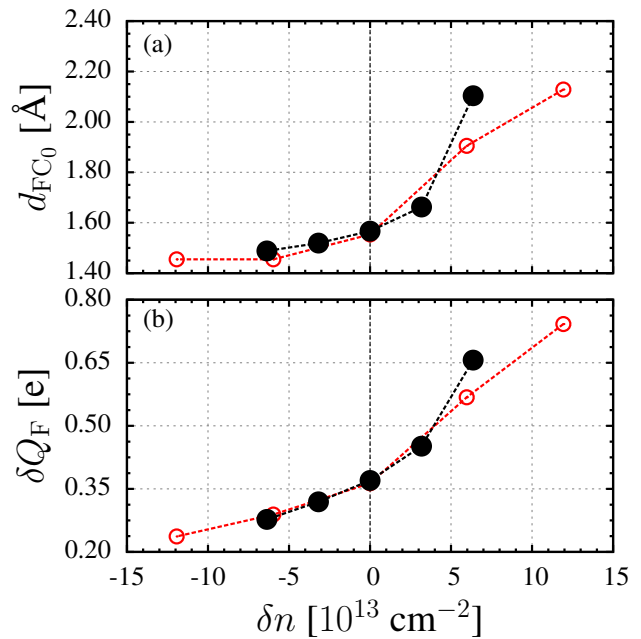}
\caption{(a) Distance between the F adatom and the C atom below it (C$_0$) as a function of doping and for two unit cell sizes: $60$C ($\bullet$) and $32$C ($\circ$). Notice that the distance $d_{\mathrm{FC}_0}$ increases as the system is charge negatively, signaling a covalent-ionic like transition. The change with the system size on the electron doped region is discussed in the text. (b) Projected charge of F for the two unit cell sizes. }
\label{dFC}
\end{figure}

To better understand what is happening with the fluorine adatom, we plot the F-C$_0$ distance ($d_{\mathrm{FC}_0}$) as a function of $\delta n$  in Fig. \ref{dFC}(a). Our DFT results show that $d_{\mathrm{FC}_0}$ significantly increases in the electron doped region. This is in agreement with the overall picture that the F atom changes its covalent bonding into a more ionic one and moves slightly apart from the graphene sheet, which recovers its planar structure.
We also include the results for a smaller unit cell (with $32$ C atoms) to show the importance of the finite size effects in the latter case, that leads to a shorter $d_{\mathrm{FC}_0}$ for the largest electron doping. The physical origin of this behavior can be understood as follows: as the graphene sheet is doped with electrons, the fluorine atom is charged and it is pushed away from graphene. This is the result of a competition between the Coulomb interaction of the charged F and  C atoms and the deformation energy of the graphene lattice.\cite{Sofo2011} Since in our DFT calculations we are not considering a single F atom but a periodic array of them separated by the size of the unit cell, there is a Coulomb repulsion between adjacent F atoms. Hence, if the unit cell is too small, that repulsion is energetically too costly and the  $d_{\mathrm{FC}_0}$  gets shorter in order to increase the covalent character of the bond and thus make the charge transfer to the F atom smaller, which in turns reduces the Coulomb energy (this is clear for the value of doping $\delta n\sim 6\times 10^{13}$ cm$^{-2}$).
This effect is absent for $\delta n<0$ as in that case the charge on the F atoms is significantly smaller. 
To support such picture, we present in Fig. \ref{dFC}(b) the charge of the F atom, projected on the $s$ and $p$ orbitals as a function of the doping. While this is not the total charge of the F atoms, it gives a reasonable idea of the above mentioned effect: the case of the $60$C units, with a smaller interaction among F atoms, allows for a larger amount of charge to be transferred to the F atoms and, consequently, the latter moves farther apart from the graphene sheet.
We note in passing that the distance between the F and the plane of the graphene sheet remains roughly constant (not shown), $\sim 2$\AA, in the entire doping region up to $\delta n\sim 5\times10^{13}$cm$^{-2}$. For larger electron doping there appears to be a sudden change where the graphene sheet flattens, the C$_0$ atom retracts back, and the F atom moves slightly apart from graphene plane, $\sim2.3$\AA---notice this is slightly larger that $d_{\mathrm{FC}_0}$.

To calculate the diffusion barrier $\Delta$ we use the NEB method to find the energy change when the F adatoms moves between two adjacent sites of the graphene lattice---this path, along the bond of a pair of C atoms, corresponds to the smaller energy barrier and therefore sets in the energy scale for the diffusion process. 
For the undoped case, we found an energy barrier of $\Delta_0\simeq0.28$meV, consisted with previous studies\cite{Wehling2009}. This value is well below the energy required to remove the F from the graphene  ($\sim1$eV), thereby ensuring that the F atoms diffuse on the graphene sheet. We also notice that the diffusion barrier for F in neutral graphene is roughly one third of the one found for O,\cite{Suarez2011} what makes the dynamics of F much faster.

\begin{figure}[tb]
\includegraphics[width=\columnwidth]{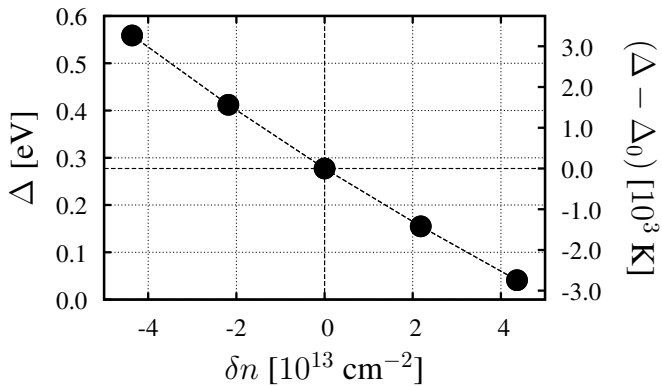}
\caption{
Diffusion barrier $\Delta$ as a function of doping. The behavior of the barrier is nearly linear and can be approximated as $\Delta(\delta n)-\Delta_0=-\alpha\delta n$, with $\alpha\simeq6\times10^{-12}$meVcm$^2$ and $\Delta_0\simeq0.28$meV.}
\label{dif}
\end{figure}

Figure \ref{dif} presents the doping dependence of the diffusion barrier, the main result of this work. In consonance with the structural change of the bonding, we found that $\Delta$ is strongly reduced when going from the hole to the electron doping region. This has a clear interpretation: while the hole doping strengthens the sp$^3$ character of the F-C bond, thereby making harder for the F atoms to jump to a nearest neighbor site, the electron doping does the contrary, creating a more ionic-like bond and a flatter graphene surface where the F atoms is more free to move. We also found that the 
distance between the F and the nearest C atoms at the intermediate or transitional point also increases with doping. The magnetic configuration, slightly more stable when the F is at the bridge position with large hole doping, have a negligible impact on the energy barrier (about $1$-$2$\%).

For low doping concentration, we find that the change of the diffusion barrier can be approximated as
\begin{equation}
\Delta(\delta n)=\Delta_0-\alpha\, \delta n\,,
\end{equation}
where  $\alpha\simeq6\times10^{-12}$meVcm$^2$ ($\simeq7\times10^{-11}$Kcm$^2$). 
Quite notably, this is about the same value of $\alpha$ that we infer  from the data of Ref. [\onlinecite{Suarez2011}] for the case of O adatoms, that might suggest a common underlying mechanism.\cite{notes2} For the F diffusion, however, the linear dependence on doping is observed around the neutrality point while in the case of O it is shifted towards the electron doping region. 

The diffusion constant $D$ depends exponentially on this barrier, $D=W \exp(-\Delta/\kt)$. By neglecting the sub-dominant change of the prefactor $W$---that depends on the local curvature of the energy landscape as a function of the F position---we found that
\begin{equation}
\frac{D(\delta n)}{D_0}=\exp{\left(\frac{\alpha \delta n}{\kt}\right)}\,.
\end{equation}
Here, $D_0=W\mathrm{e}^{-\Delta_0/\kt}$ is the neutral diffusion constant at temperature $T$. As a consequence,  even at room temperature, the diffusion constant can change by a factor of $10$ (increase or decrease depending on the character of the doping) when the doping is changed in $\sim\! 10^{13}$cm$^{-2}$.
In the low $T$ regime, say $T\sim 50$K, and for more usual densities, $\delta n\sim\pm 10^{12}$cm$^{-2}$,
the effect can still be large, leading a factor of  $10$-$20$ between hole and electron doped regions. 
We would like to point out that charge density puddles ($\delta n\sim\pm 10^{11}$cm$^{-2}$), induced by the inhomogeneous substrate potential, can lead to a significant  fluctuation of $D$ if the temperature is low enough (few kelvins). This might be relevant for the fluorine dynamics on realistic graphene samples. For example, we  speculate that the faster diffusion on electron doped regions might be more favorable for cluster formation in diluted samples that the hole doped regions.

In summary, we have showed that the diffusion of fluorine adatoms on graphene is strongly affected by doping and depends on its character (electron or hole). We believe this effect could be used, by local gating for example, to manipulate the dynamics of fluorine atoms on graphene.

We acknowledge useful discussions with J. Sofo and financial support from PICT
Bicentenario 2010-1060 and PICT 2012-379 from ANPCyT, PIP 11220080101821 and 11220110100832 from CONICET and 06/C400 and 06/C415 SeCyT-UNC. ADHN and GU acknowledges support from the ICTP associateship program. GU also thanks the Simons Foundation.


\begin{thebibliography}{36}%
\makeatletter
\providecommand \@ifxundefined [1]{%
 \@ifx{#1\undefined}
}%
\providecommand \@ifnum [1]{%
 \ifnum #1\expandafter \@firstoftwo
 \else \expandafter \@secondoftwo
 \fi
}%
\providecommand \@ifx [1]{%
 \ifx #1\expandafter \@firstoftwo
 \else \expandafter \@secondoftwo
 \fi
}%
\providecommand \natexlab [1]{#1}%
\providecommand \enquote  [1]{``#1''}%
\providecommand \bibnamefont  [1]{#1}%
\providecommand \bibfnamefont [1]{#1}%
\providecommand \citenamefont [1]{#1}%
\providecommand \href@noop [0]{\@secondoftwo}%
\providecommand \href [0]{\begingroup \@sanitize@url \@href}%
\providecommand \@href[1]{\@@startlink{#1}\@@href}%
\providecommand \@@href[1]{\endgroup#1\@@endlink}%
\providecommand \@sanitize@url [0]{\catcode `\\12\catcode `\$12\catcode
  `\&12\catcode `\#12\catcode `\^12\catcode `\_12\catcode `\%12\relax}%
\providecommand \@@startlink[1]{}%
\providecommand \@@endlink[0]{}%
\providecommand \url  [0]{\begingroup\@sanitize@url \@url }%
\providecommand \@url [1]{\endgroup\@href {#1}{\urlprefix }}%
\providecommand \urlprefix  [0]{URL }%
\providecommand \Eprint [0]{\href }%
\providecommand \doibase [0]{http://dx.doi.org/}%
\providecommand \selectlanguage [0]{\@gobble}%
\providecommand \bibinfo  [0]{\@secondoftwo}%
\providecommand \bibfield  [0]{\@secondoftwo}%
\providecommand \translation [1]{[#1]}%
\providecommand \BibitemOpen [0]{}%
\providecommand \bibitemStop [0]{}%
\providecommand \bibitemNoStop [0]{.\EOS\space}%
\providecommand \EOS [0]{\spacefactor3000\relax}%
\providecommand \BibitemShut  [1]{\csname bibitem#1\endcsname}%
\let\auto@bib@innerbib\@empty
\bibitem [{\citenamefont {Uchoa}\ \emph {et~al.}(2008)\citenamefont {Uchoa},
  \citenamefont {Kotov}, \citenamefont {Peres},\ and\ \citenamefont {{Castro
  Neto}}}]{Uchoa2008}%
  \BibitemOpen
  \bibfield  {author} {\bibinfo {author} {\bibfnamefont {B.}~\bibnamefont
  {Uchoa}}, \bibinfo {author} {\bibfnamefont {V.~N.}\ \bibnamefont {Kotov}},
  \bibinfo {author} {\bibfnamefont {N.~M.~R.}\ \bibnamefont {Peres}}, \ and\
  \bibinfo {author} {\bibfnamefont {A.~H.}\ \bibnamefont {{Castro Neto}}},\
  }\href {\doibase 10.1103/PhysRevLett.101.026805} {\bibfield  {journal}
  {\bibinfo  {journal} {Phys. Rev. Lett.}\ }\textbf {\bibinfo {volume} {101}},\
  \bibinfo {pages} {026805} (\bibinfo {year} {2008})}\BibitemShut {NoStop}%
\bibitem [{\citenamefont {Yazyev}\ and\ \citenamefont
  {Helm}(2007)}]{Yazyev2007}%
  \BibitemOpen
  \bibfield  {author} {\bibinfo {author} {\bibfnamefont {O.~V.}\ \bibnamefont
  {Yazyev}}\ and\ \bibinfo {author} {\bibfnamefont {L.}~\bibnamefont {Helm}},\
  }\href@noop {} {\bibfield  {journal} {\bibinfo  {journal} {Phys. Rev. B}\
  }\textbf {\bibinfo {volume} {75}},\ \bibinfo {pages} {125408} (\bibinfo
  {year} {2007})}\BibitemShut {NoStop}%
\bibitem [{\citenamefont {Palacios}\ \emph {et~al.}(2008)\citenamefont
  {Palacios}, \citenamefont {Fern{\'a}ndez-Rossier},\ and\ \citenamefont
  {Brey}}]{Palacios2008}%
  \BibitemOpen
  \bibfield  {author} {\bibinfo {author} {\bibfnamefont {J.~J.}\ \bibnamefont
  {Palacios}}, \bibinfo {author} {\bibfnamefont {J.}~\bibnamefont
  {Fern{\'a}ndez-Rossier}}, \ and\ \bibinfo {author} {\bibfnamefont
  {L.}~\bibnamefont {Brey}},\ }\href@noop {} {\bibfield  {journal} {\bibinfo
  {journal} {Phys. Rev. B}\ }\textbf {\bibinfo {volume} {77}},\ \bibinfo
  {pages} {195428} (\bibinfo {year} {2008})}\BibitemShut {NoStop}%
\bibitem [{\citenamefont {Yazyev}(2010)}]{Yazyev2010}%
  \BibitemOpen
  \bibfield  {author} {\bibinfo {author} {\bibfnamefont {O.~V.}\ \bibnamefont
  {Yazyev}},\ }\href@noop {} {\bibfield  {journal} {\bibinfo  {journal} {Rep.
  Prog. Phys.}\ }\textbf {\bibinfo {volume} {73}},\ \bibinfo {pages} {056501}
  (\bibinfo {year} {2010})}\BibitemShut {NoStop}%
\bibitem [{\citenamefont {Sofo}\ \emph {et~al.}(2012)\citenamefont {Sofo},
  \citenamefont {Usaj}, \citenamefont {Cornaglia}, \citenamefont {Suarez},
  \citenamefont {Hern{\'a}ndez-Nieves},\ and\ \citenamefont
  {Balseiro}}]{Sofo2012}%
  \BibitemOpen
  \bibfield  {author} {\bibinfo {author} {\bibfnamefont {J.}~\bibnamefont
  {Sofo}}, \bibinfo {author} {\bibfnamefont {G.}~\bibnamefont {Usaj}}, \bibinfo
  {author} {\bibfnamefont {P.~S.}\ \bibnamefont {Cornaglia}}, \bibinfo {author}
  {\bibfnamefont {A.}~\bibnamefont {Suarez}}, \bibinfo {author} {\bibfnamefont
  {A.~D.}\ \bibnamefont {Hern{\'a}ndez-Nieves}}, \ and\ \bibinfo {author}
  {\bibfnamefont {C.~A.}\ \bibnamefont {Balseiro}},\ }\href@noop {} {\bibfield
  {journal} {\bibinfo  {journal} {Phys. Rev. B}\ }\textbf {\bibinfo {volume}
  {85}},\ \bibinfo {pages} {115405} (\bibinfo {year} {2012})}\BibitemShut
  {NoStop}%
\bibitem [{\citenamefont {Vojta}\ and\ \citenamefont
  {Fritz}(2004)}]{Vojta2004}%
  \BibitemOpen
  \bibfield  {author} {\bibinfo {author} {\bibfnamefont {M.}~\bibnamefont
  {Vojta}}\ and\ \bibinfo {author} {\bibfnamefont {L.}~\bibnamefont {Fritz}},\
  }\href@noop {} {\bibfield  {journal} {\bibinfo  {journal} {Phys. Rev. B}\
  }\textbf {\bibinfo {volume} {70}},\ \bibinfo {pages} {094502} (\bibinfo
  {year} {2004})}\BibitemShut {NoStop}%
\bibitem [{\citenamefont {Wehling}\ \emph
  {et~al.}(2010{\natexlab{a}})\citenamefont {Wehling}, \citenamefont
  {Balatsky}, \citenamefont {Katsnelson}, \citenamefont {Lichtenstein},\ and\
  \citenamefont {Rosch}}]{Wehling2010}%
  \BibitemOpen
  \bibfield  {author} {\bibinfo {author} {\bibfnamefont {T.~O.}\ \bibnamefont
  {Wehling}}, \bibinfo {author} {\bibfnamefont {A.~V.}\ \bibnamefont
  {Balatsky}}, \bibinfo {author} {\bibfnamefont {M.~I.}\ \bibnamefont
  {Katsnelson}}, \bibinfo {author} {\bibfnamefont {A.~I.}\ \bibnamefont
  {Lichtenstein}}, \ and\ \bibinfo {author} {\bibfnamefont {A.}~\bibnamefont
  {Rosch}},\ }\href {\doibase 10.1103/PhysRevB.81.115427} {\bibfield  {journal}
  {\bibinfo  {journal} {Phys. Rev. B}\ }\textbf {\bibinfo {volume} {81}},\
  \bibinfo {pages} {115427} (\bibinfo {year} {2010}{\natexlab{a}})}\BibitemShut
  {NoStop}%
\bibitem [{\citenamefont {Cornaglia}\ \emph {et~al.}(2009)\citenamefont
  {Cornaglia}, \citenamefont {Usaj},\ and\ \citenamefont
  {Balseiro}}]{Cornaglia2009}%
  \BibitemOpen
  \bibfield  {author} {\bibinfo {author} {\bibfnamefont {P.~S.}\ \bibnamefont
  {Cornaglia}}, \bibinfo {author} {\bibfnamefont {G.}~\bibnamefont {Usaj}}, \
  and\ \bibinfo {author} {\bibfnamefont {C.~A.}\ \bibnamefont {Balseiro}},\
  }\href {\doibase 10.1103/PhysRevLett.102.046801} {\bibfield  {journal}
  {\bibinfo  {journal} {Phys. Rev. Lett.}\ }\textbf {\bibinfo {volume} {102}},\
  \bibinfo {pages} {046801} (\bibinfo {year} {2009})}\BibitemShut {NoStop}%
\bibitem [{\citenamefont {Tombros}\ \emph {et~al.}(2007)\citenamefont
  {Tombros}, \citenamefont {Jozsa}, \citenamefont {Popinciuc}, \citenamefont
  {Jonkman},\ and\ \citenamefont {van Wees}}]{Tombros2007}%
  \BibitemOpen
  \bibfield  {author} {\bibinfo {author} {\bibfnamefont {N.}~\bibnamefont
  {Tombros}}, \bibinfo {author} {\bibfnamefont {C.}~\bibnamefont {Jozsa}},
  \bibinfo {author} {\bibfnamefont {M.}~\bibnamefont {Popinciuc}}, \bibinfo
  {author} {\bibfnamefont {H.~T.}\ \bibnamefont {Jonkman}}, \ and\ \bibinfo
  {author} {\bibfnamefont {B.~J.}\ \bibnamefont {van Wees}},\ }\href {\doibase
  10.1038/nature06037, Letter} {\bibfield  {journal} {\bibinfo  {journal}
  {Nature}\ }\textbf {\bibinfo {volume} {448}},\ \bibinfo {pages} {571 }
  (\bibinfo {year} {2007})}\BibitemShut {NoStop}%
\bibitem [{\citenamefont {{Castro Neto}}\ and\ \citenamefont
  {Guinea}(2009)}]{CastroNeto2009}%
  \BibitemOpen
  \bibfield  {author} {\bibinfo {author} {\bibfnamefont {A.~H.}\ \bibnamefont
  {{Castro Neto}}}\ and\ \bibinfo {author} {\bibfnamefont {F.}~\bibnamefont
  {Guinea}},\ }\href {\doibase 10.1103/PhysRevLett.103.026804} {\bibfield
  {journal} {\bibinfo  {journal} {Phys. Rev. Lett.}\ }\textbf {\bibinfo
  {volume} {103}},\ \bibinfo {pages} {026804} (\bibinfo {year}
  {2009})}\BibitemShut {NoStop}%
\bibitem [{\citenamefont {Han}\ and\ \citenamefont {Kawakami}(2011)}]{Han2011}%
  \BibitemOpen
  \bibfield  {author} {\bibinfo {author} {\bibfnamefont {W.}~\bibnamefont
  {Han}}\ and\ \bibinfo {author} {\bibfnamefont {R.~K.}\ \bibnamefont
  {Kawakami}},\ }\href {\doibase 10.1103/PhysRevLett.107.047207} {\bibfield
  {journal} {\bibinfo  {journal} {Phys. Rev. Lett.}\ }\textbf {\bibinfo
  {volume} {107}},\ \bibinfo {pages} {047207} (\bibinfo {year}
  {2011})}\BibitemShut {NoStop}%
\bibitem [{\citenamefont {Kochan}\ \emph {et~al.}(2014)\citenamefont {Kochan},
  \citenamefont {Gmitra},\ and\ \citenamefont {Fabian}}]{Kochan2014}%
  \BibitemOpen
  \bibfield  {author} {\bibinfo {author} {\bibfnamefont {D.}~\bibnamefont
  {Kochan}}, \bibinfo {author} {\bibfnamefont {M.}~\bibnamefont {Gmitra}}, \
  and\ \bibinfo {author} {\bibfnamefont {J.}~\bibnamefont {Fabian}},\ }\href
  {\doibase 10.1103/PhysRevLett.112.116602} {\bibfield  {journal} {\bibinfo
  {journal} {Phys. Rev. Lett.}\ }\textbf {\bibinfo {volume} {112}},\ \bibinfo
  {pages} {116602} (\bibinfo {year} {2014})}\BibitemShut {NoStop}%
\bibitem [{\citenamefont {Ostrovsky}\ \emph {et~al.}(2006)\citenamefont
  {Ostrovsky}, \citenamefont {Gornyi},\ and\ \citenamefont
  {Mirlin}}]{Ostrovsky2006}%
  \BibitemOpen
  \bibfield  {author} {\bibinfo {author} {\bibfnamefont {P.~M.}\ \bibnamefont
  {Ostrovsky}}, \bibinfo {author} {\bibfnamefont {I.~V.}\ \bibnamefont
  {Gornyi}}, \ and\ \bibinfo {author} {\bibfnamefont {A.~D.}\ \bibnamefont
  {Mirlin}},\ }\href {\doibase 10.1103/PhysRevB.74.235443} {\bibfield
  {journal} {\bibinfo  {journal} {Phys. Rev. B}\ }\textbf {\bibinfo {volume}
  {74}},\ \bibinfo {pages} {235443} (\bibinfo {year} {2006})}\BibitemShut
  {NoStop}%
\bibitem [{\citenamefont {Gattenloehner}\ \emph {et~al.}(2013)\citenamefont
  {Gattenloehner}, \citenamefont {Hannes}, \citenamefont {Ostrovsky},
  \citenamefont {Gornyi}, \citenamefont {Mirlin},\ and\ \citenamefont
  {Titov}}]{Gattenloehner2013}%
  \BibitemOpen
  \bibfield  {author} {\bibinfo {author} {\bibfnamefont {S.}~\bibnamefont
  {Gattenloehner}}, \bibinfo {author} {\bibfnamefont {W.~R.}\ \bibnamefont
  {Hannes}}, \bibinfo {author} {\bibfnamefont {P.~M.}\ \bibnamefont
  {Ostrovsky}}, \bibinfo {author} {\bibfnamefont {I.~V.}\ \bibnamefont
  {Gornyi}}, \bibinfo {author} {\bibfnamefont {A.~D.}\ \bibnamefont {Mirlin}},
  \ and\ \bibinfo {author} {\bibfnamefont {M.}~\bibnamefont {Titov}},\
  }\href@noop {} {\bibfield  {journal} {\bibinfo  {journal} {arXiv.org}\ }
  (\bibinfo {year} {2013})},\ \Eprint {http://arxiv.org/abs/1306.5686v1}
  {1306.5686v1} \BibitemShut {NoStop}%
\bibitem [{\citenamefont {Roche}\ \emph {et~al.}(2012)\citenamefont {Roche},
  \citenamefont {Leconte}, \citenamefont {Ortmann}, \citenamefont {Lherbier},
  \citenamefont {Soriano},\ and\ \citenamefont {Charlier}}]{Roche2012}%
  \BibitemOpen
  \bibfield  {author} {\bibinfo {author} {\bibfnamefont {S.}~\bibnamefont
  {Roche}}, \bibinfo {author} {\bibfnamefont {N.}~\bibnamefont {Leconte}},
  \bibinfo {author} {\bibfnamefont {F.}~\bibnamefont {Ortmann}}, \bibinfo
  {author} {\bibfnamefont {A.}~\bibnamefont {Lherbier}}, \bibinfo {author}
  {\bibfnamefont {D.}~\bibnamefont {Soriano}}, \ and\ \bibinfo {author}
  {\bibfnamefont {J.-C.}\ \bibnamefont {Charlier}},\ }\href {\doibase
  http://dx.doi.org/10.1016/j.ssc.2012.04.030} {\bibfield  {journal} {\bibinfo
  {journal} {Solid State Comm.}\ }\textbf {\bibinfo {volume} {152}},\ \bibinfo
  {pages} {1404 } (\bibinfo {year} {2012})}\BibitemShut {NoStop}%
\bibitem [{\citenamefont {Hong}\ \emph {et~al.}(2011)\citenamefont {Hong},
  \citenamefont {Cheng}, \citenamefont {Herding},\ and\ \citenamefont
  {Zhu}}]{Hong2011}%
  \BibitemOpen
  \bibfield  {author} {\bibinfo {author} {\bibfnamefont {X.}~\bibnamefont
  {Hong}}, \bibinfo {author} {\bibfnamefont {S.~H.}\ \bibnamefont {Cheng}},
  \bibinfo {author} {\bibfnamefont {C.}~\bibnamefont {Herding}}, \ and\
  \bibinfo {author} {\bibfnamefont {J.}~\bibnamefont {Zhu}},\ }\href@noop {}
  {\bibfield  {journal} {\bibinfo  {journal} {Phys. Rev. B}\ }\textbf {\bibinfo
  {volume} {83}},\ \bibinfo {pages} {085410} (\bibinfo {year}
  {2011})}\BibitemShut {NoStop}%
\bibitem [{\citenamefont {Usaj}\ \emph {et~al.}(2014)\citenamefont {Usaj},
  \citenamefont {Cornaglia},\ and\ \citenamefont {Balseiro}}]{Usaj2014}%
  \BibitemOpen
  \bibfield  {author} {\bibinfo {author} {\bibfnamefont {G.}~\bibnamefont
  {Usaj}}, \bibinfo {author} {\bibfnamefont {P.~S.}\ \bibnamefont {Cornaglia}},
  \ and\ \bibinfo {author} {\bibfnamefont {C.~A.}\ \bibnamefont {Balseiro}},\
  }\href {\doibase 10.1103/PhysRevB.89.085405} {\bibfield  {journal} {\bibinfo
  {journal} {Phys. Rev. B}\ }\textbf {\bibinfo {volume} {89}},\ \bibinfo
  {pages} {085405} (\bibinfo {year} {2014})}\BibitemShut {NoStop}%
\bibitem [{\citenamefont {Cresti}\ \emph {et~al.}(2013)\citenamefont {Cresti},
  \citenamefont {Ortmann}, \citenamefont {Louvet}, \citenamefont {Van~Tuan},\
  and\ \citenamefont {Roche}}]{Cresti2013}%
  \BibitemOpen
  \bibfield  {author} {\bibinfo {author} {\bibfnamefont {A.}~\bibnamefont
  {Cresti}}, \bibinfo {author} {\bibfnamefont {F.}~\bibnamefont {Ortmann}},
  \bibinfo {author} {\bibfnamefont {T.}~\bibnamefont {Louvet}}, \bibinfo
  {author} {\bibfnamefont {D.}~\bibnamefont {Van~Tuan}}, \ and\ \bibinfo
  {author} {\bibfnamefont {S.}~\bibnamefont {Roche}},\ }\href@noop {}
  {\bibfield  {journal} {\bibinfo  {journal} {Phys. Rev. Lett.}\ }\textbf
  {\bibinfo {volume} {110}},\ \bibinfo {pages} {196601} (\bibinfo {year}
  {2013})}\BibitemShut {NoStop}%
\bibitem [{\citenamefont {Das~Sarma}\ \emph {et~al.}(2011)\citenamefont
  {Das~Sarma}, \citenamefont {Adam}, \citenamefont {Hwang},\ and\ \citenamefont
  {Rossi}}]{DasSarma2011}%
  \BibitemOpen
  \bibfield  {author} {\bibinfo {author} {\bibfnamefont {S.}~\bibnamefont
  {Das~Sarma}}, \bibinfo {author} {\bibfnamefont {S.}~\bibnamefont {Adam}},
  \bibinfo {author} {\bibfnamefont {E.~H.}\ \bibnamefont {Hwang}}, \ and\
  \bibinfo {author} {\bibfnamefont {E.}~\bibnamefont {Rossi}},\ }\href
  {\doibase 10.1103/RevModPhys.83.407} {\bibfield  {journal} {\bibinfo
  {journal} {Rev. Mod. Phys.}\ }\textbf {\bibinfo {volume} {83}},\ \bibinfo
  {pages} {407} (\bibinfo {year} {2011})}\BibitemShut {NoStop}%
\bibitem [{\citenamefont {Duplock}\ \emph {et~al.}(2004)\citenamefont
  {Duplock}, \citenamefont {Scheffler},\ and\ \citenamefont
  {Lindan}}]{Duplock2004}%
  \BibitemOpen
  \bibfield  {author} {\bibinfo {author} {\bibfnamefont {E.~J.}\ \bibnamefont
  {Duplock}}, \bibinfo {author} {\bibfnamefont {M.}~\bibnamefont {Scheffler}},
  \ and\ \bibinfo {author} {\bibfnamefont {P.~J.~D.}\ \bibnamefont {Lindan}},\
  }\href {\doibase 10.1103/PhysRevLett.92.225502} {\bibfield  {journal}
  {\bibinfo  {journal} {Phys. Rev. Lett.}\ }\textbf {\bibinfo {volume} {92}},\
  \bibinfo {pages} {225502} (\bibinfo {year} {2004})}\BibitemShut {NoStop}%
\bibitem [{\citenamefont {Meyer}\ \emph {et~al.}(2008)\citenamefont {Meyer},
  \citenamefont {Girit}, \citenamefont {Crommie},\ and\ \citenamefont
  {Zettl}}]{Meyer2008}%
  \BibitemOpen
  \bibfield  {author} {\bibinfo {author} {\bibfnamefont {J.~C.}\ \bibnamefont
  {Meyer}}, \bibinfo {author} {\bibfnamefont {C.~O.}\ \bibnamefont {Girit}},
  \bibinfo {author} {\bibfnamefont {M.~F.}\ \bibnamefont {Crommie}}, \ and\
  \bibinfo {author} {\bibfnamefont {A.}~\bibnamefont {Zettl}},\ }\href
  {\doibase 10.1038/nature07094} {\bibfield  {journal} {\bibinfo  {journal}
  {Nature}\ }\textbf {\bibinfo {volume} {454}},\ \bibinfo {pages} {319}
  (\bibinfo {year} {2008})}\BibitemShut {NoStop}%
\bibitem [{\citenamefont {Chan}\ \emph {et~al.}(2008)\citenamefont {Chan},
  \citenamefont {Neaton},\ and\ \citenamefont {Cohen}}]{Chan2008}%
  \BibitemOpen
  \bibfield  {author} {\bibinfo {author} {\bibfnamefont {K.~T.}\ \bibnamefont
  {Chan}}, \bibinfo {author} {\bibfnamefont {J.~B.}\ \bibnamefont {Neaton}}, \
  and\ \bibinfo {author} {\bibfnamefont {M.~L.}\ \bibnamefont {Cohen}},\ }\href
  {\doibase 10.1103/PhysRevB.77.235430} {\bibfield  {journal} {\bibinfo
  {journal} {Phys. Rev. B}\ }\textbf {\bibinfo {volume} {77}},\ \bibinfo
  {pages} {235430} (\bibinfo {year} {2008})}\BibitemShut {NoStop}%
\bibitem [{\citenamefont {Wehling}\ \emph {et~al.}(2009)\citenamefont
  {Wehling}, \citenamefont {Katsnelson},\ and\ \citenamefont
  {Lichtenstein}}]{Wehling2009}%
  \BibitemOpen
  \bibfield  {author} {\bibinfo {author} {\bibfnamefont {T.~O.}\ \bibnamefont
  {Wehling}}, \bibinfo {author} {\bibfnamefont {M.~I.}\ \bibnamefont
  {Katsnelson}}, \ and\ \bibinfo {author} {\bibfnamefont {A.~I.}\ \bibnamefont
  {Lichtenstein}},\ }\href {\doibase 10.1103/PhysRevB.80.085428} {\bibfield
  {journal} {\bibinfo  {journal} {Phys. Rev. B}\ }\textbf {\bibinfo {volume}
  {80}},\ \bibinfo {pages} {085428} (\bibinfo {year} {2009})}\BibitemShut
  {NoStop}%
\bibitem [{\citenamefont {Wehling}\ \emph
  {et~al.}(2010{\natexlab{b}})\citenamefont {Wehling}, \citenamefont {Dahal},
  \citenamefont {Lichtenstein}, \citenamefont {Katsnelson}, \citenamefont
  {Manoharan},\ and\ \citenamefont {Balatsky}}]{Wehling2010a}%
  \BibitemOpen
  \bibfield  {author} {\bibinfo {author} {\bibfnamefont {T.~O.}\ \bibnamefont
  {Wehling}}, \bibinfo {author} {\bibfnamefont {H.~P.}\ \bibnamefont {Dahal}},
  \bibinfo {author} {\bibfnamefont {A.~I.}\ \bibnamefont {Lichtenstein}},
  \bibinfo {author} {\bibfnamefont {M.~I.}\ \bibnamefont {Katsnelson}},
  \bibinfo {author} {\bibfnamefont {H.~C.}\ \bibnamefont {Manoharan}}, \ and\
  \bibinfo {author} {\bibfnamefont {A.~V.}\ \bibnamefont {Balatsky}},\ }\href
  {\doibase 10.1103/PhysRevB.81.085413} {\bibfield  {journal} {\bibinfo
  {journal} {Phys. Rev. B}\ }\textbf {\bibinfo {volume} {81}},\ \bibinfo
  {pages} {085413} (\bibinfo {year} {2010}{\natexlab{b}})}\BibitemShut
  {NoStop}%
\bibitem [{\citenamefont {Boukhvalov}\ and\ \citenamefont
  {Katsnelson}(2009)}]{Boukhvalov2009}%
  \BibitemOpen
  \bibfield  {author} {\bibinfo {author} {\bibfnamefont {D.~W.}\ \bibnamefont
  {Boukhvalov}}\ and\ \bibinfo {author} {\bibfnamefont {M.~I.}\ \bibnamefont
  {Katsnelson}},\ }\href {http://stacks.iop.org/0953-8984/21/i=34/a=344205}
  {\bibfield  {journal} {\bibinfo  {journal} {Journal of Physics: Condensed
  Matter}\ }\textbf {\bibinfo {volume} {21}},\ \bibinfo {pages} {344205}
  (\bibinfo {year} {2009})}\BibitemShut {NoStop}%
\bibitem [{\citenamefont {Ao}\ and\ \citenamefont {Peeters}(2010)}]{Ao2010}%
  \BibitemOpen
  \bibfield  {author} {\bibinfo {author} {\bibfnamefont {Z.~M.}\ \bibnamefont
  {Ao}}\ and\ \bibinfo {author} {\bibfnamefont {F.~M.}\ \bibnamefont
  {Peeters}},\ }\href {\doibase 10.1063/1.3456384} {\bibfield  {journal}
  {\bibinfo  {journal} {Appl. Phys. Lett.}\ }\textbf {\bibinfo {volume} {96}},\
  \bibinfo {pages} {253106} (\bibinfo {year} {2010})}\BibitemShut {NoStop}%
\bibitem [{\citenamefont {Lehtinen}\ \emph {et~al.}(2003)\citenamefont
  {Lehtinen}, \citenamefont {Foster}, \citenamefont {Ayuela}, \citenamefont
  {Krasheninnikov}, \citenamefont {Nordlund},\ and\ \citenamefont
  {Nieminen}}]{Lehtinen2003}%
  \BibitemOpen
  \bibfield  {author} {\bibinfo {author} {\bibfnamefont {P.~O.}\ \bibnamefont
  {Lehtinen}}, \bibinfo {author} {\bibfnamefont {A.~S.}\ \bibnamefont
  {Foster}}, \bibinfo {author} {\bibfnamefont {A.}~\bibnamefont {Ayuela}},
  \bibinfo {author} {\bibfnamefont {A.}~\bibnamefont {Krasheninnikov}},
  \bibinfo {author} {\bibfnamefont {K.}~\bibnamefont {Nordlund}}, \ and\
  \bibinfo {author} {\bibfnamefont {R.~M.}\ \bibnamefont {Nieminen}},\ }\href
  {\doibase 10.1103/PhysRevLett.91.017202} {\bibfield  {journal} {\bibinfo
  {journal} {Phys. Rev. Lett.}\ }\textbf {\bibinfo {volume} {91}},\ \bibinfo
  {pages} {017202} (\bibinfo {year} {2003})}\BibitemShut {NoStop}%
\bibitem [{\citenamefont {Sofo}\ \emph {et~al.}(2011)\citenamefont {Sofo},
  \citenamefont {Suarez}, \citenamefont {Usaj}, \citenamefont {Cornaglia},
  \citenamefont {Hern\'andez-Nieves},\ and\ \citenamefont
  {Balseiro}}]{Sofo2011}%
  \BibitemOpen
  \bibfield  {author} {\bibinfo {author} {\bibfnamefont {J.~O.}\ \bibnamefont
  {Sofo}}, \bibinfo {author} {\bibfnamefont {A.~M.}\ \bibnamefont {Suarez}},
  \bibinfo {author} {\bibfnamefont {G.}~\bibnamefont {Usaj}}, \bibinfo {author}
  {\bibfnamefont {P.~S.}\ \bibnamefont {Cornaglia}}, \bibinfo {author}
  {\bibfnamefont {A.~D.}\ \bibnamefont {Hern\'andez-Nieves}}, \ and\ \bibinfo
  {author} {\bibfnamefont {C.~A.}\ \bibnamefont {Balseiro}},\ }\href {\doibase
  10.1103/PhysRevB.83.081411} {\bibfield  {journal} {\bibinfo  {journal} {Phys.
  Rev. B}\ }\textbf {\bibinfo {volume} {83}},\ \bibinfo {pages} {081411}
  (\bibinfo {year} {2011})}\BibitemShut {NoStop}%
\bibitem [{\citenamefont {Chan}\ \emph {et~al.}(2011)\citenamefont {Chan},
  \citenamefont {Lee},\ and\ \citenamefont {Cohen}}]{Chan2011}%
  \BibitemOpen
  \bibfield  {author} {\bibinfo {author} {\bibfnamefont {K.~T.}\ \bibnamefont
  {Chan}}, \bibinfo {author} {\bibfnamefont {H.}~\bibnamefont {Lee}}, \ and\
  \bibinfo {author} {\bibfnamefont {M.~L.}\ \bibnamefont {Cohen}},\ }\href
  {\doibase 10.1103/PhysRevB.84.165419} {\bibfield  {journal} {\bibinfo
  {journal} {Phys. Rev. B}\ }\textbf {\bibinfo {volume} {84}},\ \bibinfo
  {pages} {165419} (\bibinfo {year} {2011})}\BibitemShut {NoStop}%
\bibitem [{\citenamefont {Suarez}\ \emph {et~al.}(2011)\citenamefont {Suarez},
  \citenamefont {Radovic}, \citenamefont {Bar-Ziv},\ and\ \citenamefont
  {Sofo}}]{Suarez2011}%
  \BibitemOpen
  \bibfield  {author} {\bibinfo {author} {\bibfnamefont {A.~M.}\ \bibnamefont
  {Suarez}}, \bibinfo {author} {\bibfnamefont {L.~R.}\ \bibnamefont {Radovic}},
  \bibinfo {author} {\bibfnamefont {E.}~\bibnamefont {Bar-Ziv}}, \ and\
  \bibinfo {author} {\bibfnamefont {J.~O.}\ \bibnamefont {Sofo}},\ }\href
  {\doibase 10.1103/PhysRevLett.106.146802} {\bibfield  {journal} {\bibinfo
  {journal} {Phys. Rev. Lett.}\ }\textbf {\bibinfo {volume} {106}},\ \bibinfo
  {pages} {146802} (\bibinfo {year} {2011})}\BibitemShut {NoStop}%
\bibitem [{\citenamefont {Giannozzi}\ \emph {et~al.}(2009)\citenamefont
  {Giannozzi}, \citenamefont {Baroni}, \citenamefont {Bonini}, \citenamefont
  {Calandra}, \citenamefont {Car}, \citenamefont {Cavazzoni}, \citenamefont
  {Ceresoli}, \citenamefont {Chiarotti}, \citenamefont {Cococcioni},
  \citenamefont {Dabo}, \citenamefont {Corso}, \citenamefont {de~Gironcoli},
  \citenamefont {Fabris}, \citenamefont {Fratesi}, \citenamefont {Gebauer},
  \citenamefont {Gerstmann}, \citenamefont {Gougoussis}, \citenamefont
  {Kokalj}, \citenamefont {Lazzeri}, \citenamefont {Martin-Samos},
  \citenamefont {Marzari}, \citenamefont {Mauri}, \citenamefont {Mazzarello},
  \citenamefont {Paolini}, \citenamefont {Pasquarello}, \citenamefont
  {Paulatto}, \citenamefont {Sbraccia}, \citenamefont {Scandolo}, \citenamefont
  {Sclauzero}, \citenamefont {Seitsonen}, \citenamefont {Smogunov},
  \citenamefont {Umari},\ and\ \citenamefont {Wentzcovitch}}]{QE}%
  \BibitemOpen
  \bibfield  {author} {\bibinfo {author} {\bibfnamefont {P.}~\bibnamefont
  {Giannozzi}}, \bibinfo {author} {\bibfnamefont {S.}~\bibnamefont {Baroni}},
  \bibinfo {author} {\bibfnamefont {N.}~\bibnamefont {Bonini}}, \bibinfo
  {author} {\bibfnamefont {M.}~\bibnamefont {Calandra}}, \bibinfo {author}
  {\bibfnamefont {R.}~\bibnamefont {Car}}, \bibinfo {author} {\bibfnamefont
  {C.}~\bibnamefont {Cavazzoni}}, \bibinfo {author} {\bibfnamefont
  {D.}~\bibnamefont {Ceresoli}}, \bibinfo {author} {\bibfnamefont {G.~L.}\
  \bibnamefont {Chiarotti}}, \bibinfo {author} {\bibfnamefont {M.}~\bibnamefont
  {Cococcioni}}, \bibinfo {author} {\bibfnamefont {I.}~\bibnamefont {Dabo}},
  \bibinfo {author} {\bibfnamefont {A.~D.}\ \bibnamefont {Corso}}, \bibinfo
  {author} {\bibfnamefont {S.}~\bibnamefont {de~Gironcoli}}, \bibinfo {author}
  {\bibfnamefont {S.}~\bibnamefont {Fabris}}, \bibinfo {author} {\bibfnamefont
  {G.}~\bibnamefont {Fratesi}}, \bibinfo {author} {\bibfnamefont
  {R.}~\bibnamefont {Gebauer}}, \bibinfo {author} {\bibfnamefont
  {U.}~\bibnamefont {Gerstmann}}, \bibinfo {author} {\bibfnamefont
  {C.}~\bibnamefont {Gougoussis}}, \bibinfo {author} {\bibfnamefont
  {A.}~\bibnamefont {Kokalj}}, \bibinfo {author} {\bibfnamefont
  {M.}~\bibnamefont {Lazzeri}}, \bibinfo {author} {\bibfnamefont
  {L.}~\bibnamefont {Martin-Samos}}, \bibinfo {author} {\bibfnamefont
  {N.}~\bibnamefont {Marzari}}, \bibinfo {author} {\bibfnamefont
  {F.}~\bibnamefont {Mauri}}, \bibinfo {author} {\bibfnamefont
  {R.}~\bibnamefont {Mazzarello}}, \bibinfo {author} {\bibfnamefont
  {S.}~\bibnamefont {Paolini}}, \bibinfo {author} {\bibfnamefont
  {A.}~\bibnamefont {Pasquarello}}, \bibinfo {author} {\bibfnamefont
  {L.}~\bibnamefont {Paulatto}}, \bibinfo {author} {\bibfnamefont
  {C.}~\bibnamefont {Sbraccia}}, \bibinfo {author} {\bibfnamefont
  {S.}~\bibnamefont {Scandolo}}, \bibinfo {author} {\bibfnamefont
  {G.}~\bibnamefont {Sclauzero}}, \bibinfo {author} {\bibfnamefont {A.~P.}\
  \bibnamefont {Seitsonen}}, \bibinfo {author} {\bibfnamefont {A.}~\bibnamefont
  {Smogunov}}, \bibinfo {author} {\bibfnamefont {P.}~\bibnamefont {Umari}}, \
  and\ \bibinfo {author} {\bibfnamefont {R.~M.}\ \bibnamefont {Wentzcovitch}},\
  }\href {http://www.quantum-espresso.org} {\bibfield  {journal} {\bibinfo
  {journal} {Journal of Physics: Condensed Matter}\ }\textbf {\bibinfo {volume}
  {21}},\ \bibinfo {pages} {395502} (\bibinfo {year} {2009})}\BibitemShut
  {NoStop}%
\bibitem [{\citenamefont {Perdew}\ \emph {et~al.}(1996)\citenamefont {Perdew},
  \citenamefont {Burke},\ and\ \citenamefont {Ernzerhof}}]{pbe}%
  \BibitemOpen
  \bibfield  {author} {\bibinfo {author} {\bibfnamefont {J.~P.}\ \bibnamefont
  {Perdew}}, \bibinfo {author} {\bibfnamefont {K.}~\bibnamefont {Burke}}, \
  and\ \bibinfo {author} {\bibfnamefont {M.}~\bibnamefont {Ernzerhof}},\
  }\href@noop {} {\bibfield  {journal} {\bibinfo  {journal} {Phys. Rev. Lett.}\
  }\textbf {\bibinfo {volume} {77}},\ \bibinfo {pages} {3865} (\bibinfo {year}
  {1996})}\BibitemShut {NoStop}%
\bibitem [{\citenamefont {Bl\"ochl}(1994)}]{Blochl1994}%
  \BibitemOpen
  \bibfield  {author} {\bibinfo {author} {\bibfnamefont {P.~E.}\ \bibnamefont
  {Bl\"ochl}},\ }\href {\doibase 10.1103/PhysRevB.50.17953} {\bibfield
  {journal} {\bibinfo  {journal} {Phys. Rev. B}\ }\textbf {\bibinfo {volume}
  {50}},\ \bibinfo {pages} {17953} (\bibinfo {year} {1994})}\BibitemShut
  {NoStop}%
\bibitem [{\citenamefont {Neugebauer}\ and\ \citenamefont
  {Scheffler}(1992)}]{Neugebauer1992}%
  \BibitemOpen
  \bibfield  {author} {\bibinfo {author} {\bibfnamefont {J.}~\bibnamefont
  {Neugebauer}}\ and\ \bibinfo {author} {\bibfnamefont {M.}~\bibnamefont
  {Scheffler}},\ }\href {\doibase 10.1103/PhysRevB.46.16067} {\bibfield
  {journal} {\bibinfo  {journal} {Phys. Rev. B}\ }\textbf {\bibinfo {volume}
  {46}},\ \bibinfo {pages} {16067} (\bibinfo {year} {1992})}\BibitemShut
  {NoStop}%
\bibitem [{not({\natexlab{a}})}]{notes1}%
  \BibitemOpen
  \href@noop {} {} \ \bibinfo {note} {The Quantum Expresso
  package allows for the used of fractional filling. Present exchange
  correlation functionals only give approximate results for fractional
  occupations, however, it gives an estimate of the overall trend supported by
  the cases of integer filling.}\BibitemShut {Stop}%
\bibitem [{not({\natexlab{b}})}]{notes2}%
  \BibitemOpen
  \href@noop {} {} \ \bibinfo {note} {It should be mentioned
  that while the bonding of the adatom is different in the two cases, the diffusion process in both
  cases involve the on top and brigde positions as initial and transition states (though with their role exchanged for the O). Also, in the two cases there is a charge transfer process to the adatom involved.}\BibitemShut {Stop}%
\end{thebibliography}

%

\end{document}